# Impact of mechanical disturbances to the placement of reference light in Fourier transform interferometers

muqian wen

## Abstract

Fourier transform interferometers usually need a dedicated reference light to precisely track optical path difference changes. There are a few different ways to incorporate the reference light into the optical system. This paper conducted experiments to give a test of how different ways of incorporating reference light are affected by mechanical disturbances. It found that placing the reference light in different path than the test light can be very sensitive to mechanical disturbance errors while propagating the reference light together with the test light can be very robust against mechanical vibrations of optical components.

## 1. Introduction

Fourier transform interferometer is an instrument that in its typical form uses a moving mirror in an interferometer to generate interference signal for measurement purposes. It is typically used in spectroscopy for studying material absorption or emission etcetera such as the Fourier transform infrared spectroscopy common in chemistry [1] and it is also used in many other optical fields such as in nonlinear optics for femtosecond laser pulse characterization etcetera [2,3,4]. Optical experiments involving Fourier transform interferometers and alike typically need to use a dedicated reference light such as a helium neon laser to track real time optical path difference changes generated by the movement of translation devices precisely. The measured result is then used to either adjust the translation stage position in real time or to facilitate processing measurement data later. The measurement needs to be very accurate in those situations. There are in general three different ways to incorporate the reference light beam in the experiment designs. The first way is to merge reference light source into a single light beam with the measurement light to travel in the optical systems together [5,6,7]. The second way is to make the reference light beam travel side by side in parallel with the measurement light beam to go through the same optical elements but not to overlap with each other [8,9,10]. The third way is to have the reference beam and measurement beam travel in separate paths through different optical elements [11,12,13]. However, there are no clearcut advantages between these reference incorporation methods with each method having different shortcomings. In the first method of co-propagating the reference light with measurement light, the drawback is that it can be difficult to separate reference light and measurement light from each other. In the second method of parallel propagating reference light with measurement light, the drawback is that the optical elements such as the mirrors and beam splitters in the setup may not be big enough to accommodate two beams simultaneously without overlapping. This is especially true in the situations of miniaturing Fourier transform spectrometers [14,15] which has become an important research topic in recent years. Another limitation is that the optical elements must be able to cover both wavelengths of the reference and test light sources in parallel propagation. The third method of propagating reference light in separate optical paths may seem most enticing as it offers most freedom in the arrangements of optical elements, but its drawback is that it may be very difficult to make sure that the measured optical path position is accurate due to potential vibrations of optical elements during scan.

Given that there are quite a few ways to incorporate reference light, finding out the optimal design for an experiment may not be trivial. Mechanical disturbance is an important factor that can significantly influence design choices. It would be worthwhile to study the conditions affecting design choices to help researchers save time in finding optimal ways to incorporate reference light in their experiments. Although there have been some studies on how mechanical disturbance induced optical misalignments would affect spectral results in Fourier transform spectrometers [16,17], as far as this study knows there has been no study on how mechanical vibrations would affect the accuracy of measured optical path positions in the different ways of incorporating reference source and how this would affect the experiment designs. Studying vibration induced error will not only be relevant to Fourier transform spectroscopy experiments but also to other optical experiments that also use translation stages such as the spectral shearing interferometry [3,4].

Thus, this paper will try to conduct experiments to give a quantitative assessment on how the three different ways of incorporating reference lights in Fourier transform spectrometers will be affected by mechanical vibrations. It will build several Fourier transform spectrometers with different ways of incorporating the reference light using common off shelf optical instruments to quantitatively compare how different ways of incorporating reference light are affected by mechanical vibrations of optical elements. It will use one single mode helium neon laser source as reference source to measure the spectrum of another multimode helium neon laser. As a bonus, the spectrum of helium neon laser is also obtained at a resolution of about $1/2.5$ m$^{-1}$ which normally could not be obtained with regular lower resolution spectrometers. It is hoped that this study can be useful to other researchers using reference lights to measure optical path positions in their experiments.

## 2. Experiment

Figure 1 shows the experiment designs of this paper. The first design shown is a co-propagating symmetrical Mach–Zehnder interferometer type of Fourier transform spectrometer by using some hollow cube corner retroreflectors. Several polarizers were used to polarize the reference and test laser source in mutually perpendicular directions so that they can be separated in the photodetectors. Mach–Zehnder type design is used here instead of the conventional Michelson design to avoid having some light reflecting back into the lasers. In the conventional Michelson type of design, half of the input light is reflected back into the light source by the mirrors and this can cause significant errors in spectral result [18]. This problem is especially critical to this study because this experiment is using both helium neon lasers with same wavelength for reference and test light sources. If some of their lights are getting into each other, then it can no longer be assumed that these two lasers are independent light sources, and the results obtained in this way will no longer be guaranteed to be valid. This experiment also used a symmetrical design where optical path lengths of both arms are changed symmetrically by a single translation stage. Such symmetrical designs can not only double the total optical path difference but also have the benefit of having better tolerance of optical misalignments [19]. This experiment would use the simplest nonuniform sampling data acquisition schemes where the data were acquired at a fixed sampling rate in time while the translation stage moves continuously with varying speeds. The acquired nonuniform data can then be processed by either nonuniform fast Fourier transform or by resampling into equidistant data by interpolation [20].

For the second experiment design of parallel propagating the light beams, this experiment moved the position of the test laser by some distance so that it will not overlap with the reference beam. Both the vertical heights and horizontal positions of the test laser beam can be changed. Polarizers are still used so that the obtained spectra can be more suitable to be compared with the co-propagation one and to remove residual stray lights. The distance of the test and reference beam is about several millimeters apart which is just enough to make them not overlap with each other.

Finally, the separate propagation experiment design is a totally different construct although the equipment used is still the same. This experiment simply built two independent classical Michelson interferometer type Fourier transform spectrometers but let them share the same translation stage. The Mach–Zehnder interferometer is not necessary here because the two lasers will not be interfering with each other while the Michelson design will allow retaining the same scanning length as the two previous experiment setups. Actually, this experiment was conducted before the other two experiments. It was the undesirable results from this experiment that inspired this study to further investigate and better understand mechanical vibration induced errors.

At last, an additional experiment was conducted to obtain the full spectrum of the test helium neon laser source by discarding the polarizers and moving the test light beam further away from reference beam which can be achieved by swapping the position of the test laser with one of the photodetectors.

The optical alignments in these experiments have been adjusted to best effort as the equipment allowed. The reference laser used is a Thorlabs HRS015 frequency stabilized 632.8 nm helium neon laser. This laser is already polarized so there is no need to add an additional polarizer. Its polarization direction is set to be perpendicular to the reflection plane in this experiment. This laser also is an excellent single mode laser. The test laser used here is a Melles Griot 05-LHR-191 nonpolarized 632.8 nm helium neon laser. The polarizers used are Newport 5511 general purpose sheet polarizer. The beam splitter used is Thorlabs BS013 non-polarizing cube beam splitter. The mirrors used are generic optical mirrors with spring containing mounts to allow tuning their angles. The two big and small hollow cube corner retroreflectors are from Melles Griot PLX and Newport respectively. The retroreflectors serve to fold the beams to increase the optical path length multiple times. The translation stage used is a Physik Instrumente M-531.DDX 306 mm linear translation stage. The photodetectors used were two identical 340 nm-1100 nm Thorlabs PDA100A-EC Switchable Gain Detectors. A National Instrument USB-6341 X series multichannel data acquisition device was used to acquire the interferogram samples. The sampling rate was set to the maximum of 250 K/s per channel.

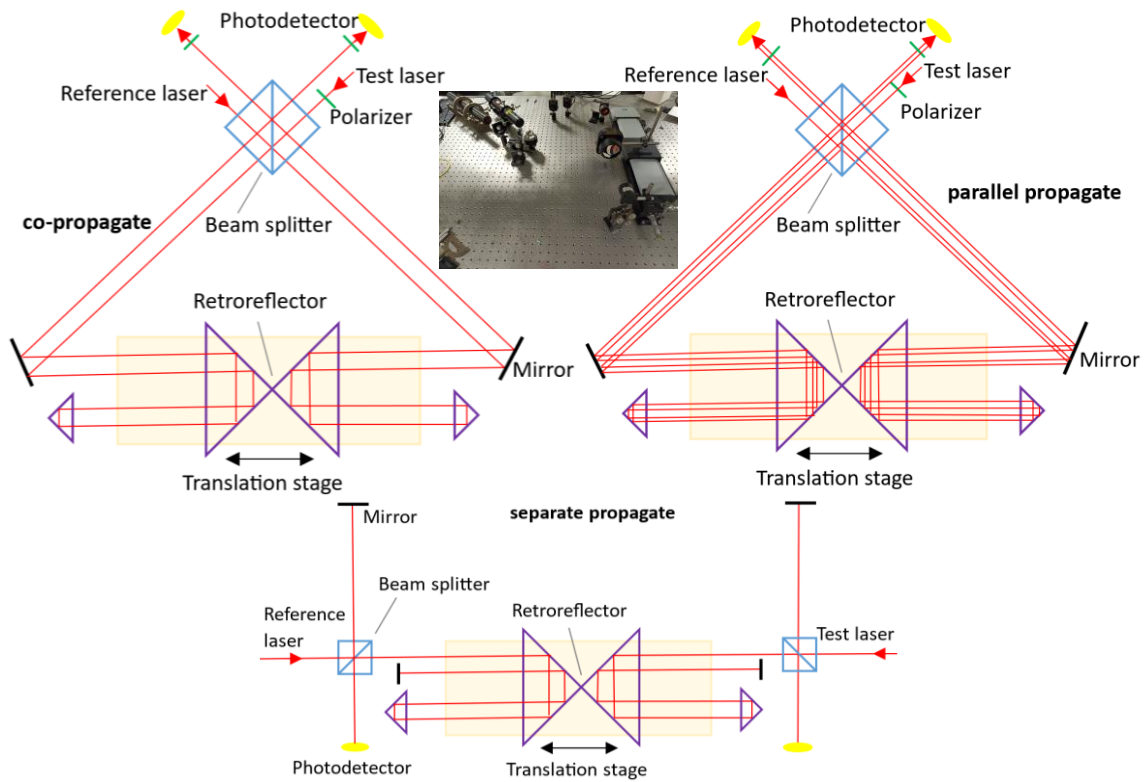

Figure 1. Designs of the experiments of co-propagation (top left), parallel propagation (top right) and separate propagation (bottom). A picture of the parallel propagation setup (top center) is also provided.

## 3. Result

Figure 2 shows an example of the interferograms obtained with co-propagating experiment and the resulting calculated test helium neon laser spectrum (in one polarization direction only). The optical path difference is estimated from the reference interferogram. The time domain spectrum of the reference and test interferograms are also shown to demonstrate how much the translation stage moving speed would vary during the scan.

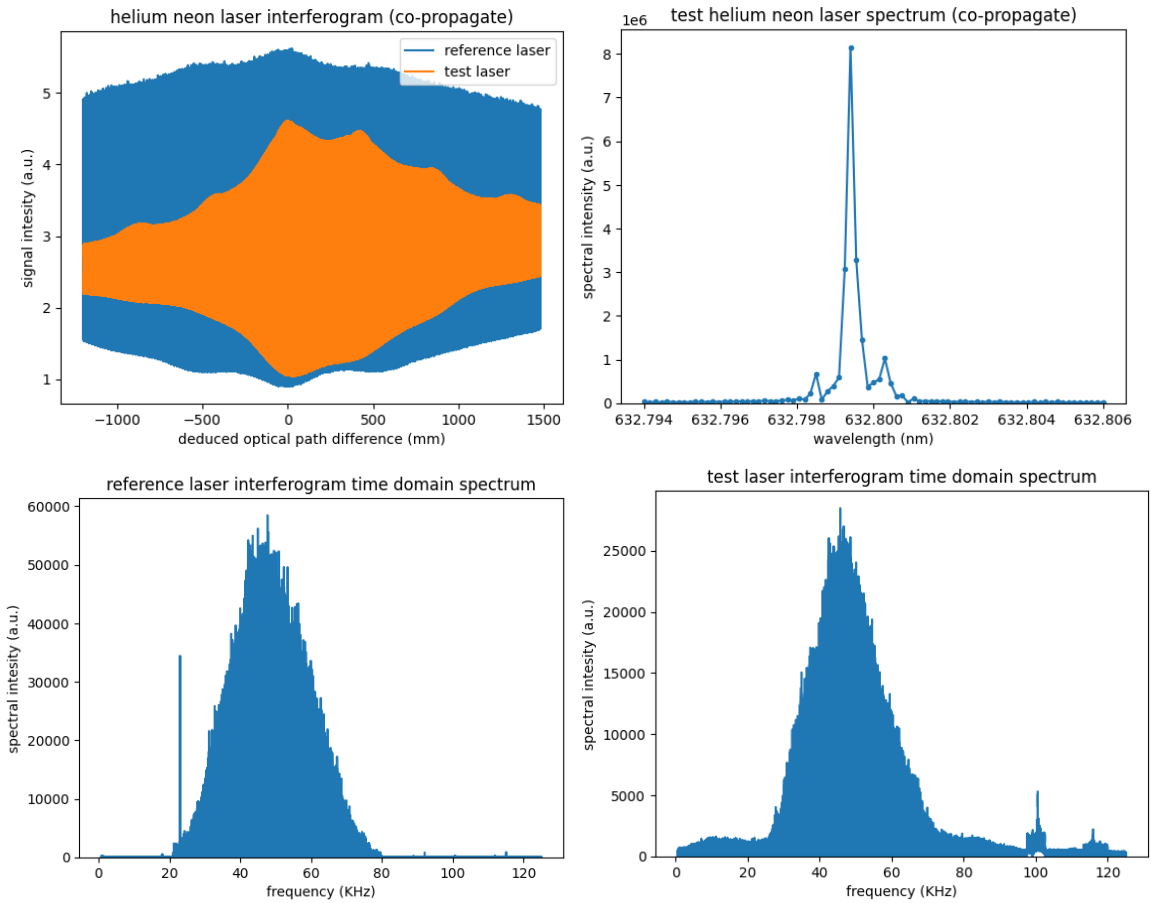

*Figure 2. The obtained interferograms (top left) in co-propagating configuration and the corresponding calculated polarized test helium neon laser spectrum (top right). The time domain spectrum of reference interferogram (bottom left) and test interferogram (bottom right) is also shown.*

Figure 3 shows an example of interferograms of parallel propagation and the resulting test laser spectrum. This spectrum cannot be directly compared with the co-propagation one because the tube of this laser had been rotated during setup changes and the polarization state of this laser will also change from run to run in this experiment.

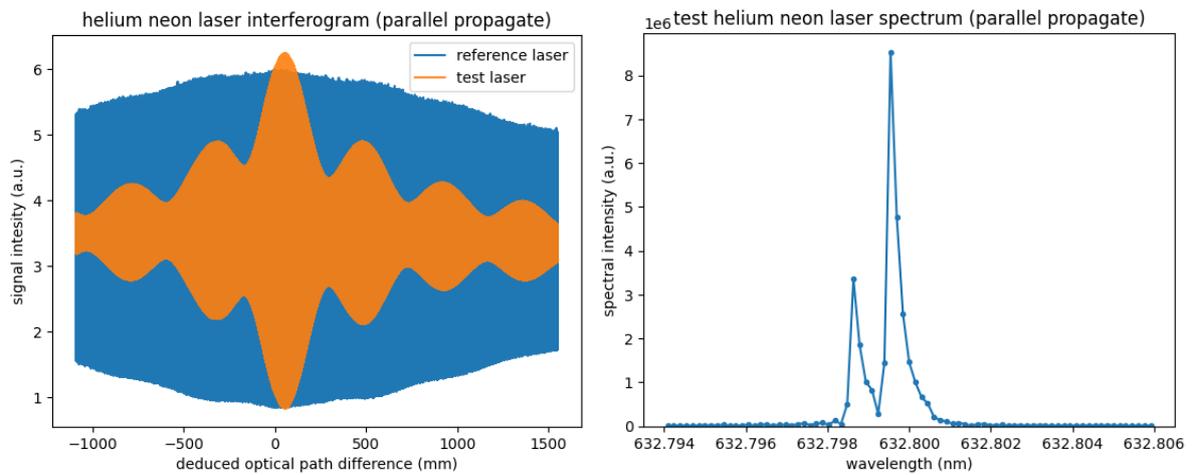

*Figure 3. The obtained interferograms (left) in parallel propagating experiment and the corresponding calculated polarized test helium neon laser spectrum (right).*

Figure 4 shows a 1 mm long segment of the interferograms of the separate propagation experiment, the corresponding test laser spectrum and time domain spectrum of test laser interferogram, as well as the difference in optical path difference changes calculated independently from the two lasers. In this case of only 1 mm long length of interferogram segment both helium neon lasers can be considered as monochromatic and thus both lasers can be used to determine the optical path difference changes accurately. It can be seen from the calculated spectrum that the sampling point positions determined in this way are too erroneous. Since a very short 1 mm segment is already not working, it would not be worthwhile to show the full length interferograms further. The calculated difference in scanning length changes shows that it can vary by as much as 1000 nm within this 1 mm segment which can explain why this setup would fail to obtain correct spectrum of the test laser.

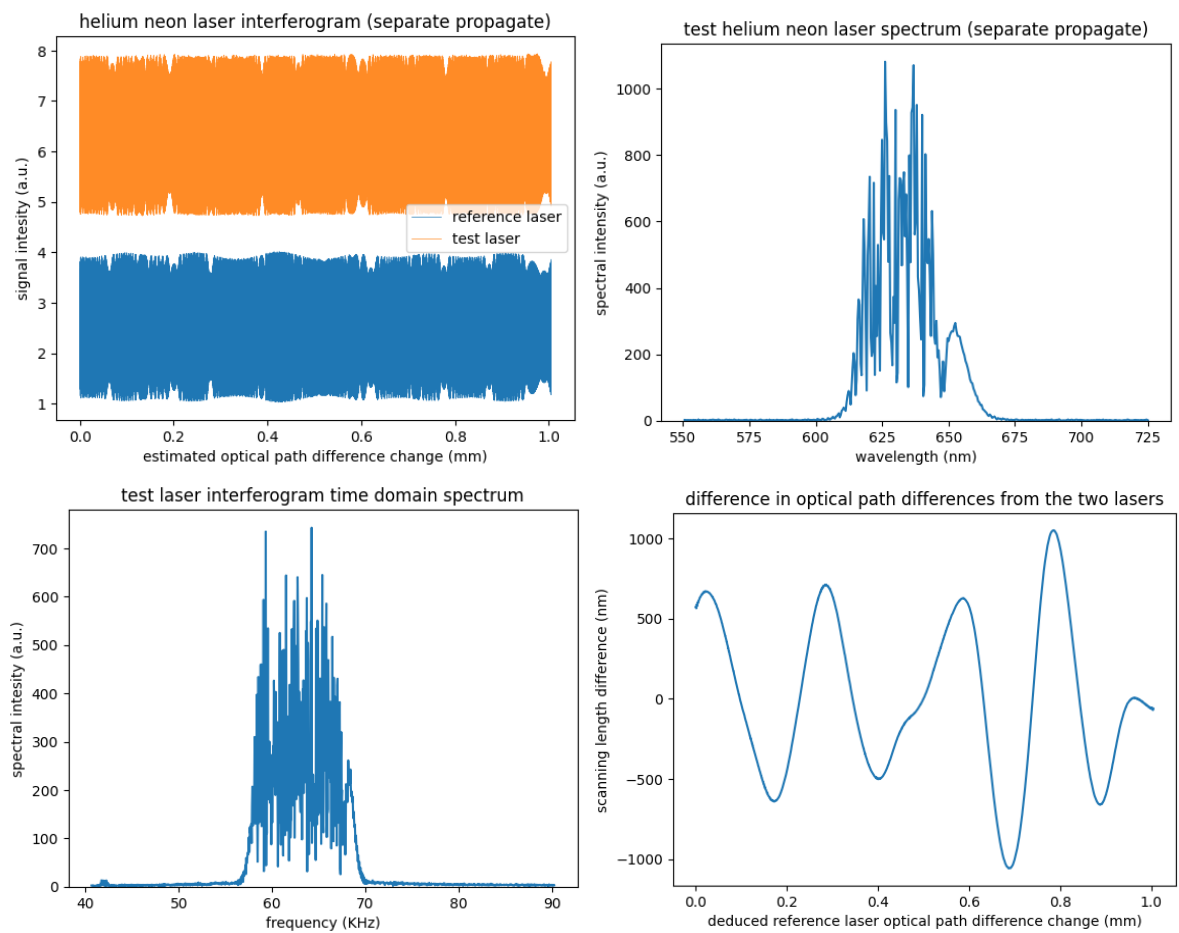

*Figure 4. A 1 mm segment of the interferograms (top left) of the separate propagation experiment, the calculated test laser spectrum (top right), the time domain spectrum of the test laser interferogram (bottom left), and the difference in scanning lengths calculated from the two helium neon lasers respectively (bottom right).*

Finally, an attempt was made to measure the full spectrum of the test helium neon laser source by parallel propagating the light without use of polarizers and Figure 5 shows an example of the result. It can be seen from the figure that there are four peaks in the test laser spectrum and the resolution of this spectrometer is just at the limit to successfully resolving them.

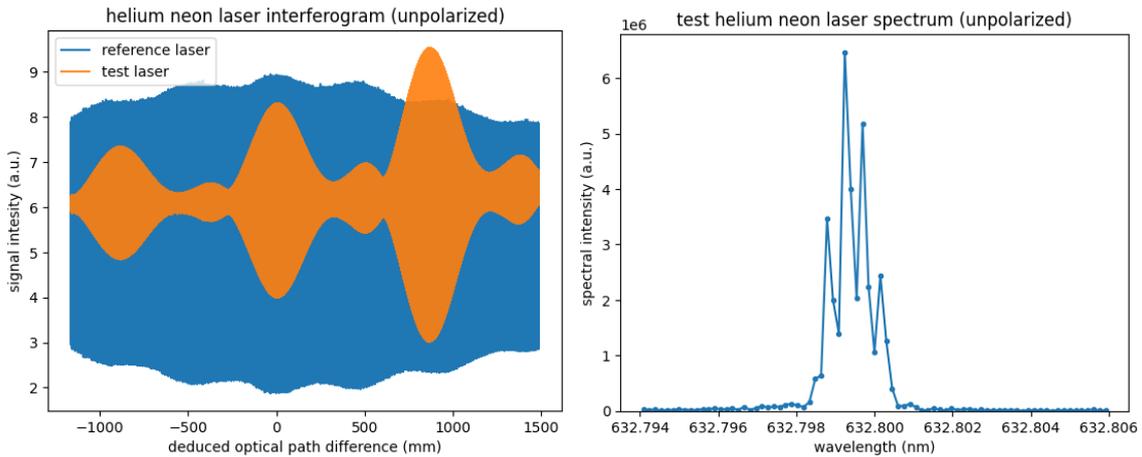

*Figure 5. Interferograms of reference and test lasers without polarization filters obtained in parallel propagating configuration (left) and the corresponding calculated test laser spectrum (right).*

## 4. Discussion

The separated propagation design experiment result shows that this method fails to work completely because even a 1 mm long segment of scan cannot work. This must be due to the mechanical vibrations of the optical elements because this experiment design has used the same optical elements and has the same maximum scanning length as the other two designs, yet it would produce dramatically different results than the others. The difference in scanning length changes calculated independently from the interferograms of the two lasers shown in figure 4 indicates that the optical elements may vibrate by as much as about 1000 nm at relatively low frequencies during the 1 mm scan length. This magnitude of vibration would certainly be too much for ultraviolet to near infrared spectrum light sources. For mid to far infrared lights this may be more acceptable but likely will still cause a lot of spectral errors. Overall, this experiment has proved that it would be very difficult to use only regular optical elements to realize separate propagating designs. But separated propagation design is certainly achievable because there have been many such examples existing in the literature [11,12,13]. And there are situations where separate propagation design must be used such as when the reference light can cause damage to sensitive photodetectors used for the test light beams [11]. There are measures to overcome mechanical disturbance issues. For example, there are optical tables with active vibration isolation mechanisms that can maintain the stability of mounted optical elements and there exists mirror mounts designed to be more resistant to external disturbances. The vibrations from translation stage motors can also be eliminated by using step scanning mechanism. However, when all these measures are required, it may become just more convenient and cost efficient to use co-propagating or parallel propagating designs instead. This study did not have the resources to test what measures would be necessary to successfully implement separate propagating designs though.

Theoretically mechanical disturbances can also cause errors in parallel propagating designs. When a mirror tilts, the reference beam path and test beam path will have different changes in path length. This path length differences error will be proportional to how far these two beams have been spaced apart. This error can be calculated by the following mathematical relationship. Supposed that in the classical Michelson interferometer design the tilting of mirror will cause the light beam to move by a distance at the photodetector, then the position error will be this distance multiply by the division between the separation distance of the two beams and the distance of the mirror to the photodetector. Thus, the error in parallel propagation will be highly dependent on how much the mirror tilts or the equivalent during the scan. The result from this experiment shows that the vibrational error if exist will be well within acceptable levels. Another set of data acquired immediately after this result without any changes to setup also showed very similar spectrum profiles thus proving that the vibrational errors in this experiment must be very small. The reference beam and test beam are far enough to not overlap with each other. Given that the vibrational error level is small, the outcome will likely be not much different if this study had access to bigger optical equipment to create larger separation distances. The design of this experiment has one big benefit in that it ensures the returning light beams from both arms will always be parallel to each other at the receiving end of photodetectors. This property can greatly simplify the analysis of the experiment results. Because in the classical Michelson interferometer design, the tilting of mirror will also cause the returning beams to become not parallel. If errors were found in the spectral result, then it will be less certain about whether to attribute the errors to changes of beam angles or changes in path lengths.

In theory co-propagating design should be mostly immune to mechanical disturbances because the reference and test beams will experience the same disturbances and thus cancel each other out. The experiment result in this study agrees well with this expectation. The resulting test laser spectrum has been confined to a very narrow range. Thus, even if the profile feature within this narrow range had been affected by mechanical errors, the spectrum will still have a very good effective resolution. The time domain discrete Fourier transform spectrum of reference interferogram and test interferogram also showed some interesting differences. The reference time domain spectrum is confined within a relatively narrow range while the test laser time domain spectrum has base values throughout the spectrum range. This is probably another indirect evidence that the reference laser is different from the test laser.

Finally, this study implemented a parallel propagating design without polarizers to try to measure the full spectrum of the test laser. The result agrees with the polarized spectrum results obtained before. It shows that this laser has about 4 output modes. The relative strength of these modes will vary from run to run between different sets of data, but such behavior is expected for a multimode laser. The resolution of this spectrometer with an about 2.4-meter scanning length is still not high enough to fully resolve these laser modes. This experiment has demonstrated that Fourier transform spectrometer is an excellent tool to measure the spectrum of long coherence length lasers.

The reference single mode helium neon laser used in this study should be assumed to be an ideal monochromatic source which can be indirectly verified by the fact that the amplitudes of its interferograms do not vary much during the scan. The main reason for the variations in amplitude is that the laser beam diameter will naturally expand as it propagates in space. It can also be seen from the reference interferograms that there exist some minor misalignments which is normal because it will not be possible to achieve perfect alignment by hand. In the experiments of this study the translation stage is mounted on the same optical table with other optical components. The translation stage will generate vibrations during movement and these vibrations will certainly transmit to other optical elements. The mirror mounts used are just regular mounts with spring connected flexible joints to allow adjusting mirror angles. The beam splitter is also mounted on similar flexible mounts. This flexibility may make them more susceptible to vibrations. The translation stage is not the only source of mechanical vibrations because when the translation stage is not moving, it can still be observed with naked eye that there are flickering of light intensities at the photodetectors at low frequencies of about several hertz. Thus, some of the vibrations must come from the ambient environment. In addition, simply touching the translation stage can also cause flickering of light which indicates that the bending of rigid structures can also be potentially an issue. The vibration of optical components can be indirectly seen from the experiment results in those many tiny irregular amplitude fluctuations in the interferogram which are likely due to ephemeral displacements of the light beams.

Overall, this study has shown that mechanical disturbances can cause unacceptable levels of errors in Fourier transform spectroscopy experiments. This implies that any optical experiments that need translation stages for precise optical path changes must use a dedicated reference source to track the position changes accurately. Many translation stage instruments have built-in mechanisms to record the translation position. But these built-in measurements cannot reflect the vibrations of the mounted optical components no matter how precise they are. In fact, it has been attempted before in the literature to use the translation stage recorded position data to process acquired interferogram data but was found not to work due to vibrational errors [11].

## 5. Conclusion

This paper has built several high-resolution Fourier transform spectrometers to help shed light on how different ways of incorporating the reference light used for tracking optical path difference changes precisely will be affected by mechanical disturbances in the accuracy of their measurements. It found that propagating the reference light in different paths using only regular optical equipment without any mechanism to prevent vibration of the optical elements will suffer heavily from mechanical vibration errors and failed to obtain usable results while co-propagating and parallel propagating reference light implementations using the same equipment will be largely immune from mechanical vibration errors. Thus, separate propagating design should be implemented with extra care or avoided when mechanical vibrations can become a concern. This paper hopes that this study can be useful to experimenters of Fourier transform spectroscopy or other optical experiments that need to use a reference light to track movement of translation stages precisely.